\documentclass[a4paper]{article}

\usepackage{INTERSPEECH2021}

\title{A HIGHER ORDER MINKOWSKI LOSS FOR IMPROVED PREDICTION ABILITY OF ACOUSTIC MODEL IN ASR}
\name{Vishwanath Pratap  Singh$^1$, Shakti P. Rath$^2$, and Abhishek Pandey$^1$}
\address{
  $^1$Samsung R\&D Institute India - Bangalore\\
  $^2$Reverie Language Technologies, India}
\email{vp.singh@samsung.com, shakti.rath@reverieinc.com,  and abhi3.pandey@samsung.com}

\begin{document}

\maketitle
\begin{abstract}
  Conventional automatic speech recognition (ASR) system uses second-order minkowski loss during inference time which is suboptimal as it incorporates only first order statistics in posterior estimation \cite{bishop}. In this paper we have proposed higher order minkowski loss (${4}^{th}$ Order and ${6}^{th}$ Order) during inference time, without any changes during training time. The main contribution of the paper is to show that higher order loss uses higher order statistics in posterior estimation, which improves the prediction ability of acoustic model in ASR system. We have shown mathematically that posterior probability obtained due to higher order loss is function of second order posterior and thus the method can be incorporated in standard ASR system in an easy manner. It is to be noted that all changes are proposed during test(inference) time, we do not make any change in any training pipeline. Multiple baseline systems namely, TDNN1, TDNN2, DNN and LSTM are developed to verify the improvement incurred due to higher order minkowski loss. All experiments are conducted on LibriSpeech dataset and performance metrics are word error rate (WER) on "dev-clean", "test-clean", "dev-other" and "test-other" datasets.
\end{abstract}

\section{Introduction}

In conventional acoustic modeling, in the inference time, the optimum classes are obtained by passing the input features through the network, which predicts the class-conditional (posterior) probabilities of the probability density function (pdf) classes given the input features. Afterwards Viterbi decoding is applied on the class-conditional probabilities to get the ASR output \cite{viterbi3}. Use of the class-conditional probabilities during inference time intuitively make sense. In section \ref{sec:2ndOrder}, through sound theoretical development, we substantiate the intuition to show that class-conditional probabilities are indeed obtained if  ${2}^{nd}$ order Minkowski loss (i.e., mean-square loss) is used as the optimum expected loss. As pointed out in \cite{bishop} and explained in section \ref{sec:2ndOrder}, the 
${2}^{nd}$  order Minkowski loss prediction model uses only  ${1}^{st}$ order statistics to predict the class-conditionals and ignores all higher order statistics, which is suboptimal. 

Higher Order Statistics provide supplementary information to the estimation. These statistics are very useful in problems where either non-Gaussianity, nonminimum phase, colored noise, or nonlinearities are important and must be accounted for. Adding higher order statistics makes the estimation robust \cite{hos}. Impact of higher-order loss approximation for deep neural networks interpretation is studied for image classification task \cite{icmlr}. It is found that higher order approximation helps when classes are more and the prediction probability is not close to one, which makes it's application suitable for acoustic model in ASR system where output classes are thousands of context dependent states.    

In this work we show that higher-order Minkowski loss (e.g.${4}^{th}$/${6}^{th}$ order expected loss) leads to better prediction ability of the acoustic model in ASR system, which makes use of both ${1}^{st}$ order as well as higher-order statistics. Specifically, the class-conditionals are modified to generate more optimum class-predictions. Afterwards, Viterbi decoding is applied to the modified class-conditionals.  We emphasize that all changes are proposed during test(inference) time, we do not make any change in any training pipeline. It is observed that modifying the  ${2}^{nd}$ order loss with higher -order minkowski loss to incorporate higher order statistics in prediction indeed improves the prediction ability of Acoustic Model in ASR, which leads to significant  reduction in word error rate. We emphasize that there is no change in the acoustic model training pipeline. The changes are made only in the decoding/inference time. The same Acoustic model is decode with ${2}^{nd}$ order minkowski loss and higher order minkowski loss and results are compared. Following the norms accepted in ASR community, in this paper we show the results on 960 hours of Librispeech dataset for Large Vocabulary Continuous Speech Recognition (LVCSR) Task. The extended work on bigger dataset is deferred to fututre work . 

 TDNN-HMM \cite{TDNN1}, DNN-HMM \cite{dnn1,dnn2} and LSTM-HMM  \cite{lstm1} decoded with second order minkowski loss is used as a baseline. In a TDNN, different layers or sets of layers can act on different time scales. As such, it can be seen as a type of convolutional neural network (CNN) operating over the time dimension. In a TDNN model, the first few layers look at smaller time scales and produce more abstract higher level features. The later layers take larger time windows over these abstract features as input. During training and inference, the sequence of input features are repeatedly shifted in time and fed to the model, producing another sequence as output. This architecture reduces the amount of computation required, as compared with a fully connected network \cite{TDNN1, TDNN2}. On the other hand long-short term model (LSTM) was proposed as an alternative to conventional RNN \cite{rnn} for speech recognition. It helps to solve the issues faced by RNN by replacing the recurrent neurons in RNN with memory cells \cite{lstm1}. In order to reduce the computational complexity of the network, the Long Short-Term Memory Projected (LSTMP)  architecture is proposed by Sak et al. \cite{lstm2}.

This paper is organised as follows. In Section~\ref{sec:2ndOrder}, we present limitation of 2nd order minkowski loss. Section~\ref{sec:4thOrder} and Section~\ref{sec:6thOrder},  explains higher order minkowski loss and their advantage over ${2}^{nd}$ order minkowski loss. Section~\ref{sec:issues} explains the issues with application of ${3}^{rd}$ and ${5}^{th}$ order minkowski loss. Data set and experimental setup are explained in Section~\ref{sec:data} and Section~\ref{sec:exp}, respectively. Conclusion with future works are provided in Section 8.

\section{Review of Second Order Minkowski Loss}\label{sec:2ndOrder}
The inference (decoding) stage in conventional ASR system consists of choosing a specific estimate $y(x)$ of the value of target ($t$) for each input $x$. Suppose that in doing so, we incur a loss $L(y(x),t)$. Then average or expected loss, as worked out in \cite{bishop}, is given by:
\begin{align}
\mathbf{E}[L] = \sum_{t=1}^{N}  \int L(y(x),t)p(x,t)dx 
\label{eq_minkowski1}
\end{align}
Where, $N$ is number of context dependent classes in acoustic model and $p(x,t)$ is join probability over input $x$ and target $t$ and ${E}[L]$ is expected loss. In case of second order minkowski loss $L(y(x),t)$ is defined as $\{y(x) - t\}^{2}$. Thus eq. \ref{eq_minkowski1} becomes:
\begin{align}
\mathbf{E}[L] = \sum_{t=1}^{N}  \int \{y(x) - t\}^{2}p(x,t)dx 
\label{eq_minkowski2}
\end{align}
Our goal is to choose $y(x)$ so as to minimize $E[L]$. Minimizing eq. \ref{eq_minkowski2} using differential method:
\begin{align}
\frac{\delta \mathbf{E}[L]}{\delta \mathbf{y(x)}} = 2\sum_{t=1}^{N}  \{y(x) - t\}p(x,t)=0
\label{eq_minkowski3}
\end{align}
Solving for y(x), and using the sum and product rules of probability, we obtain:
\begin{align}
\mathbf{y(x) } = \frac{\sum_{t=1}^{N} \mathbf{t p(x,t)} } {\mathbf{p(x)}} = \mathbf{\sum_{t=1}^{N} \mathbf{t p(t \| x)}}=\mathbf{E}_{t}[t\|x]
\label{eq_minkowski4}
\end{align}
and if targets are discrete such as probability density function (pdf) classed in acoustic model then,
\begin{align}
\mathbf{E}_{t}[t\|x] = \mu_{i}
\label{eq_minkowski5}
\end{align}
Thus, the ${2}^{nd}$ order Minkowski loss prediction model uses only ${1}^{st}$ order statistics to predict the posteriors and ignores all higher order statistics in estimation, which is suboptimal.

\section{Fourth Order Minkowski Loss} \label{sec:4thOrder}
\begin{figure}[t]
	\centering
	\includegraphics[width=8.8cm]{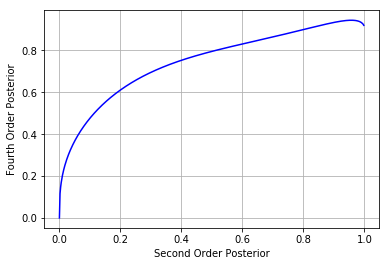}
	\caption{Correspondence between 2nd Order and 4th Order Posteriors.}
	\label{fig:fig1}
\end{figure}

Expected Loss for fourth order minkowski can be obtained by substituting loss function $L( y(x), t)$ in eq. \ref{eq_minkowski1} with $\{y(x) - t\}^{4}$ and can be defined as:
\begin{align}
\mathbf{E}[L] = \sum_{t=1}^{N}  \int \{y(x) - t\}^{4}p(x,t)dx 
\label{eq_minkowski7}
\end{align}
Expected loss defined in eq. \ref{eq_minkowski7} can be minimized by finding root of: 
\begin{align}
\frac{\delta \mathbf{E}[L]}{\delta \mathbf{y(x)}} = 4\sum_{t=1}^{N}  \{y(x) - t\}^{3}p(x,t)
\label{eq_minkowski8}
\end{align}
After expanding the cubic term and  representing $y(x)$ as $y$ we obtain:
\begin{align}
\frac{\delta \mathbf{E}[L]}{\delta \mathbf{y(x)}} = 4\sum_{t=1}^{N} ( y^{3} - 3 t {y}^2 + 3 {t}^{2} y - {t}^{3}) p(x,t)
\label{eq_minkowski21}
\end{align}
since $t$ is target which takes the binary value, i.e., $t\epsilon\{0,1\}$ we obtain:
\begin{align}
\mathbf{\sum_{t=1}^{N} \mathbf{t p(t \| x)}}=\mathbf{\sum_{t=1}^{N} \mathbf{ {t}^{n} p(t \| x)}} = \mathbf{E}_{t}[t\|x] = \mu
\label{eq_minkowski22}
\end{align}
After integrating the eq. \ref{eq_minkowski21} over discrete target and ignoring the constant multiplication term,
\begin{align}
\frac{\delta \mathbf{E}[L]}{\delta \mathbf{y(x)}} = y^{3} - 3\mu {y}^2 + 3 \mu y -\mu 
\label{eq_minkowski9}
\end{align}
The root of the function obtained in  eq. \ref{eq_minkowski9} is the modified posterior  $\mu^{4th}_{i}$, which is the function of actual posterior $\mu_{i}$. Correspondence between posterior obtained after ${2}^{nd}$ order minkowski loss  $\mu_{i}$ and posteriors obtained after 4th order minkowski loss $\mu^{4th}_{i}$ is shown in figure \ref{fig:fig1}.

\begin{table}[t]
	\caption{Improvement using 4th Order Minkowski Loss and 6th Order Minkowski Loss over TDNN-HMM baseline (TDNN) trained on 960 hours combined "train-clean" and "train-others" dataset}
	\label{tab:table2}
	\centering
	\begin{tabular}{c|c|c|c|c}
		\textbf{Test Case}      & \textbf{LM} & \textbf{TDNN}   & \textbf{4th Order} & \textbf{6th Order} \\ \hline
		dev-clean   & 3-gram small   & 7.26  & 6.91  & 6.83\\
		test-clean  & 3-gram small  & 7.58  & 7.22 & 7.13\\	
		dev-others  & 3-gram small  & 17.06  & 15.89 & 15.74 \\
		test-others  &  3-gram small  & 17.57 & 16.49 & 16.34 \\\hline
		dev-clean   & 3-gram medium   & 6.42  & 6.16 & 6.11\\
		test-clean  & 3-gram medium  & 6.78  & 6.44 & 6.32\\
		dev-others  & 3-gram medium  & 15.71  & 14.98 & 14.76\\
		test-others  &  3-gram medium  & 16.15 & 16.32 & 16.24\\ \hline
		dev-clean   & 3-gram large   & 5.04  & 4.78 & 4.71\\	
		test-clean  & 3-gram large & 5.64   & 5.39 & 5.28 \\
		dev-others  & 3-gram large  &  13.22 & 12.61 & 12.55\\
		test-others  &  3-gram large  & 13.64 & 12.96 & 12.84\\	\hline
	\end{tabular}
\end{table}

\begin{table}[t]
	\caption{Improvement using 4th Order Minkowski Loss and 6th Order Minkowski Loss over DNN-HMM baseline (DNN) trained on 960 hours combined "train-clean" and "train-others" dataset}
	\label{tab:table3}
	\centering
	\begin{tabular}{c|c|c|c|c}
		\textbf{Test Case}      & \textbf{LM} & \textbf{DNN}   & \textbf{4th Order} & \textbf{6th Order } \\ \hline
		dev-clean   & 3-gram small   & 7.61  & 7.08 & 6.97\\
		test-clean  & 3-gram small  & 7.93  & 7.42 & 7.26 \\	
		dev-others  & 3-gram small  & 17.89  & 16.76  & 16.55 \\
		test-others  &  3-gram small  & 18.45 & 17.34 & 17.19 \\ \hline
		dev-clean   & 3-gram medium   & 6.72  & 6.34 & 6.26 \\
		test-clean  & 3-gram medium  & 7.04  & 6.70  & 6.54 \\
		dev-others  & 3-gram medium  & 16.48  &  15.61 & 15.44\\
		test-others  &  3-gram medium  & 16.95 & 16.03 & 15.98 \\ \hline
		dev-clean   & 3-gram large   & 5.17  & 4.88 & 4.79\\	
		test-clean  & 3-gram large & 5.88   & 5.56 & 5.51\\
		dev-others  & 3-gram large  &  13.88 & 13.17 & 13.06 \\
		test-others  &  3-gram large  & 14.25 & 13.43 & 13.38 \\ \hline	
	\end{tabular}
\end{table}

\begin{table}[t]
	\caption{Improvement using 4th Order Minkowski Loss and 6th Order Minkowski Loss over LSTM-HMM baseline (LSTMP) trained on 960 hours combined "train-clean" and "train-others" dataset}
	\label{tab:table4}
	\centering
	\begin{tabular}{c|c|c|c|c}
		\textbf{Test Case}      & \textbf{LM} & \textbf{LSTMP}   & \textbf{4th Order} & \textbf{6th Order } \\ \hline
		dev-clean   & 3-gram small   & 6.18  & 5.83  & 6.67\\
		test-clean  & 3-gram small  & 6.43  & 5.93 & 5.72 \\	
		dev-others  & 3-gram small  & 15.56  & 14.52 & 14.13 \\
		test-others  &  3-gram small  & 16.07 & 15.01 & 14.72 \\\hline
		dev-clean   & 3-gram medium   & 5.47  & 5.03 & 4.87 \\
		test-clean  & 3-gram medium  & 5.58  & 5.14 & 5.01\\
		dev-others  & 3-gram medium  & 14.61  & 13.72 & 13.37\\
		test-others  &  3-gram medium  & 15.01 & 14.11 & 13.78\\ \hline
		dev-clean   & 3-gram large   & 4.11  & 3.79 & 3.65\\	
		test-clean  & 3-gram large & 4.52   & 4.18 & 4.07 \\
		dev-others  & 3-gram large  &  12.03 & 11.29 & 11.03\\
		test-others  &  3-gram large  & 12.35 & 11.49 & 11.21\\	\hline
	\end{tabular}
\end{table}

\begin{figure}[t]
	\centering
	\includegraphics[width=8.8cm]{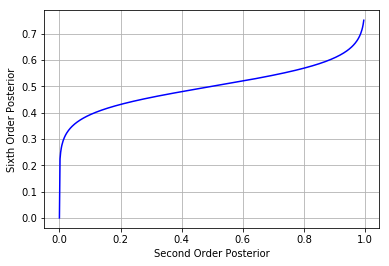}
	\caption{Correspondence between 2nd Order and 6th Order Posteriors.}
	\label{fig:fig2}
\end{figure}

\section{Sixth Order Minkowski Loss} \label{sec:6thOrder}
Expected Loss for sixth order minkowski can be obtained by substituting loss function $L(y(x),t)$ in eq. \ref{eq_minkowski1} with $\{y(x) - t\}^{6}$ and can be defined as:
\begin{align}
\mathbf{E}[L] = \sum_{t=1}^{N}  \int \{y(x) - t\}^{6}p(x,t)dx 
\label{eq_minkowski10}
\end{align}
Expected loss defined in eq. \ref{eq_minkowski7} can be minimized by finding root of: 
\begin{align}
\frac{\delta \mathbf{E}[L]}{\delta \mathbf{y(x)}} = 6\sum_{t=1}^{N}  \{y(x) - t\}^{5}p(x,t)
\label{eq_minkowski11}
\end{align}
After integrating the eq. \ref{eq_minkowski11} over discrete target, as illustrated in Section~\ref{sec:2ndOrder} and Section~\ref{sec:4thOrder}, we obtain:
\begin{align}
\frac{\delta \mathbf{E}[L]}{\delta \mathbf{y(x)}} = y^{5} - 5\mu {y}^4 + 10 \mu {y}^{3} -10\mu {y}^{2} + 5\mu y - \mu
\label{eq_minkowski12}
\end{align}
The root of the function obtained in  eq. \ref{eq_minkowski9} is the modified posterior $\mu^{6th}_{i}$ which is the function of actual posterior $\mu_{i}$. Roots of the function obtained in eq. \ref{eq_minkowski12} are obtained using newtons method \cite{newton}. Correspondence between posterior obtained after ${2}^{nd}$ order minkowski loss $\mu_{i}$ and posteriors obtaianed ${6}^{th}$ order minkowski loss $\mu^{6th}_{i}$, is shown in figure \ref{fig:fig2}.

\section{Issues with Third Order and Fifth Order Minkowski Loss} \label{sec:issues}
Expected loss for ${3}^{rd}$ Order and ${5}^{th}$ Order minkoski loss can be defined by modifying the eq.\ref{eq_minkowski1} and given by:
\begin{align}
\mathbf{E}[L] = \sum_{t=1}^{N}  \int \{y(x) - t\}^{3}p(x,t)dx 
\label{eq_minkowski13}
\end{align}
\begin{align}
\mathbf{E}[L] = \sum_{t=1}^{N}  \int \{y(x) - t\}^{5}p(x,t)dx 
\label{eq_minkowski14}
\end{align}
As illustrated in section \ref{sec:4thOrder} and section \ref{sec:6thOrder} the differentiation of ${3}^{rd}$ Order and ${5}^{th}$ Order loss can be obtained and given by eq. \ref{eq_minkowski15} and eq. \ref{eq_minkowski16}, respectively:
\begin{align}
\frac{\delta \mathbf{E}[L]}{\delta \mathbf{y(x)}} = y^{2} -\mu {y} + \mu
\label{eq_minkowski15}
\end{align}
\begin{align}
\frac{\delta \mathbf{E}[L]}{\delta \mathbf{y(x)}} = y^{4} - 4\mu y^{3} + 6 \mu y^{2} -4 \mu y + \mu
\label{eq_minkowski16}
\end{align}
when we analyzed the roots of eq. \ref{eq_minkowski15} and eq. \ref{eq_minkowski16} they turned out to be complex numbers ($i.e., a + ib$) thus can't be used as a probability which is always real and between 0 to 1. Hence, these ${3}^{rd}$ and ${5}^{th}$  order minkowski loss can't be used during inference time.

\section{Data Description} \label{sec:data}
LibriSpeech data set is used to train and evaluate the proposed methods.
LibriSpeech training dataset consist of about 960 hours of
read audio books. The training set of corpus is divided into two subsets,
approximately 460 hours "train-clean" and 500 hours "train-others", respectively \cite{libri}. Speed perturbation \cite{sp} of the training data to obtain 3 copies of the training data corresponding to speed perturbations of 0.9, 1.0 and 1.1 were created. The dev and test sets were split into simple (”clean”) and harder (”other”).  Test sets "test-clean" and "test-others" contains $5.4$ and $5.1$  hours of data, respectively. Similarly, dev sets "dev-clean" and "dev-others" contains $5.4$ and $5.3$  hours of data, respectively.

\section{Experimental Analysis and Results} \label{sec:exp}
All experiments are conducted using the Kaldi toolkit \cite{kaldi}. Librispeech data set, as explained in \ref{sec:data}, is used to train and validate all experiments. The GMM-HMM was trained on Mel frequency cepstral coefficients (MFCC) using LDA+MLLT \cite{mllt,stc} transformation. A GMM-HMM system was trained in the LDA+STC space. There were $3952$ senones in total. Afterwards discriminative training was applied using boosted MMI \cite{bmmi}. All the models are initialized using layerwise discriminative pre-training, followed by  cross-entropy (CE) \cite{ce} fine-tuning. A WFST-based decoder \cite{viterbi3} has been used. All experiments are conducted using either DNN, TDNN or LSTM based acoustic model to evaluate the performance of our proposed methods. Truncated back propagation through time (BPTT) \cite{bptt} with fixed truncation length of $20$ is used for network training. The performance metrics are word error rate(WER). Three different language model are used for decoding the proposed acoustic model. In the results shown in  \ref{tab:table2, table3, table4}, 3-gram samll language model(LM) is the pruned LM, which is used for lattice generation. 3-gram medium is slightly less pruned 3-gram LM and the 3-gram large LM is the full, non-pruned 3-gram LM. The relation between $2^{nd}$ order posterior, $4^{th}$ order posterior and $6^th$ order posterior is shown in figure \ref{fig:fig1}  and figure \ref{fig:fig2}, respectively. It is observed that, slope is sharp for small posteriors which indicates higher order statistics is more significant when the posteriors are weak while there is linear relation between $2^{nd}$ order posteriors and Higher order posteriors which indicates the minimal role of higher order statistics in case of strong posteriors. It is also observed that adding $3^{rd}$ Order Statistics ($4^{th}$ order loss) gives significant improvement and adding higher order statistics up to $5^{th}$ order ($6^{th}$ Order Loss) giver little  improvement on top of $6^{th}$ Order Loss which Indicates that $3^{rd}$ Order statistics itself captures the most of the non-linearity and skewness. 

\subsection{TDNN-HMM Baseline} \label{sub-sec:tdnn}
Time-Delay Neural Network (TDNN) \cite{TDNN1} is used as the second baseline Acoustic Model. The use of TDNN is motivated by the fact that it performs at par with recurrent neural network based acoustic models \cite{TDNN1,TDNN2} while still being able to be trained in parallel due to the absence of recurrent connections. They are successful in learning short-term and long-term dependencies in the input signal using only short-term acoustic features. The configuration is, splices together frames t - 2 through t + 2 at the input layer (which we could write as context     $\{ - 2,  - 1, 0, 1, 2\}$
; and then at $7$ hidden layers we splice frames at offsets
$\{ 1, 1\}$ ,  $\{ 3, 2\}$ , $\{ 5, 3\}$ , $\{  7, 4\}$ , $\{  9, 5\}$ , $\{  11, 7\}$ and $\{ 13, 9\}$.  
Finally, there is an output layer with cross entropy loss across $3953$ context-dependent phone states.  TDNN-HMM baseline is trained on 960 hours of Librispeech data set/ For brevity we will refer to this architectures as TDNN and \ref{tab:table2}.

\subsection{DNN-HMM Baseline} \label{sub-sec:dnn}
The second baseline system is a a feed-forward 7 layers DNN-HMM network. The input to DNN-HMM comprised $143$ dimensional features generated by mean-normalized MFCCs with context expansion of $\pm 5$, followed by LDA normalization. The number of hidden units per layer is 1024. Single DNN-HMM baselines trained on 960 hours of complete Librispeech data set is used in this experiment. For brevity we will refer to this architecture as DNN1.  Results are presented in table \ref{tab:table3}.

\subsection{LSTM-HMM Baseline} \label{sub-sec:lstm}
The third baseline system is a 5 layers LSTM with projection (LSTMP) network \cite{lstm2}. Each layers contained 1024 memory cells and a recurrent projection layer of size $256$. The input to model comprised $65$ dimensional features, obtained by splicing the mean-normalized MFCCs using a context of $\pm3$, further normalized by LDA transformation without dimensionality reduction. 40 frames are used as left context to predict the output label for the current frame. Output HMM state label is delayed by 5 time steps to see the information from future frames while predicting for thes current frame \cite{lstm2}.  Results are presented in table \ref{tab:table4}.

\subsection{Improvement using Fourth Order Minkowski Loss }
Posteriors of above 3 baselines, as explained in section \ref{sub-sec:tdnn}, \ref{sub-sec:dnn} and \ref{sub-sec:lstm} are modified using the ${4}^{th}$ order minkowski loss \ref{sec:4thOrder} and passed to veterbi decoder. Proposed method is evaluated on 4 test namely "test-clean", "test-others", "dev-clean" and "dev-others". Results are presented in table \ref{tab:table2}, \ref{tab:table3} and \ref{tab:table4}. It is observed that adding higher order statistics indeed improves the accuracy. The best performance using ${4}^{th}$ order minkowski loss  gives relative reduction word error rate as follows, 6.8\%(On "dev-clean" dataset using large 3-gram arpa) on top of TDNN1 baseline, 6.7\%(On "dev-other" dataset using 3-gram small arpa ) on top of TDNN2 baseline, 7.1\% (On "dev-clean" data set using 3-gram small arpa) on top of DNN baseline and 7.9\% (On "dev-"clean dataset using 3-gram large arpa) on top of LSTM-HMM baseline.

\subsection{Improvement using Sixth Order Minkowski Loss }
Motivated by the performance of the  ${4}^{th}$ order minkowski loss which uses up to third order statistics in posterior estimation, we experimented with ${6}^{th}$ order minkowski loss which uses upto ${5}^{th}$ order statistics. Results are presented in table \ref{tab:table1} , table \ref{tab:table2} and table \ref{tab:table3}. ${6}^{th}$  Order minkowski loss gives further reduction in word error rate which amounts to, 7.5\%(On "dev-other" dataset using 3-gram small arpa) on top of TDNN1 baseline, 8.1\%(On "dev-other" dataset using 3-gram small) on top of TDNN2 baseline, 7.9\%(On "test-clean" dataset using 3-gram large) on top of DNN baseline and 9.8\% (On "dev-"clean dataset using 3-gram large arpa) on top of LSTM-HMM baseline,  relative reduction is WER. It is to be noted here that the  ${6}^{th}$  Order minkowski loss giver little improvement on top of ${4}^{th}$ order minkowski loss which indicates that adding higher order statistics further doesn't help much in prediction ability of acoustic model in ASR.

\section{Conclusions} \label{sec:conclusion}
In this paper we explore the advantage of higher order minkowski losss over ${2}^{nd}$ minkowski loss in ASR for large vocabulary continuous speech recognition task during inference time. We show that the higher order minkoski loss uses the higher order statistics in posterior estimation which improves the prediction ability of acoustic model in ASR system, which results into significant reduction in word error rate. The experiments are conducted with TDNN, DNN and LSTM and results are obtained with combination of weak and strong language model, we observe the consistent improvement across all combinations.  We have shown mathematically that posterior probability obtained due to higher order loss is function of second order posterior and  thus the acoustic model output can be modified to obtain higher order posterior in an easy manner. Moreover, there is no change in acoustic model training pipeline.

\newpage
\cleardoublepage

\end{document}